\documentclass[10pt,letterpaper]{article}
\usepackage{opex3}
\usepackage{setspace}
\usepackage{amsmath,amssymb,graphicx}
\usepackage{cite}

\begin{document}

\title{Interplay between multiple scattering, emission, and absorption of light in the phosphor of a white light-emitting diode}

\author{V.Y.F. Leung$^{1}$, A. Lagendijk$^{1}$, T.W. Tukker$^{2}$, A.P. Mosk$^{1}$, W.L. IJzerman$^{3}$, and W.L. Vos$^{1}$}

\address{$^{1}$Complex Photonic Systems (COPS), MESA+ Institute for Nanotechnology, University of Twente, P.O. Box 217, Enschede 7500 AE, The Netherlands}
\address{$^{2}$Philips Research, High Tech Campus 34, Eindhoven 5656 AE, The Netherlands}
\address{$^{3}$Philips Lighting, High Tech Campus 44, Eindhoven 5656 AE, The Netherlands}
\email{y.f.v.leung@utwente.nl} 
\homepage{http://cops.nano-cops.com/} 


\begin{abstract}
  We study light transport in phosphor plates of white light-emitting diodes (LEDs).  We measure the broadband diffuse transmission through phosphor plates of varying YAG:Ce$^{3+}$ density. We distinguish the spectral ranges where absorption, scattering, and re-emission dominate.  Using diffusion theory, we derive the transport and absorption mean free paths from first principles.  We find that both transport and absorption mean free paths are on the order of the plate thickness. This means that phosphors in commercial LEDs operate well within an intriguing albedo range around 0.7.  We discuss how salient parameters that can be derived from first principles control the optical properties of a white LED.
\end{abstract}




\section{Introduction}

White-light emitting diodes (LEDs) are rapidly replacing traditional lighting as an energy efficient form of illumination.  The most common white LED today consists of a blue semiconductor diode combined with luminescent phosphors that partially convert the blue light to yellow and red~\cite{Sch06, Kra07, Bec08}. This energy transfer occurs in a structurally complex environment.  The phosphor layer is intentionally engineered to incorporate air inclusions and material scatterers.  As a result light is scattered several times within the layer, making light transport a diffusive phenomenon. Broadband light diffusion contributes to the achievement of an even lighting without hot spots or angular color distribution, essential factors for many lighting applications.  In addition, the recycling of photons by multiple scattering allows thinner phosphor layers to be used.  This increases cost efficiency and reduces environmental impact.

The combination of diffusive light scattering and energy conversion in phosphor-converted LEDs is most often described by ray-tracing and Monte Carlo techniques~\cite{Gil96, Ben04, Som09, Liu10, Tuk10}.  Being numerical simulations, these are not capable of offering fundamental physical insights with predictive power, which could guide the evolution of efficient phosphor layers with improved optical properties.  In addition, simulations can be computationally demanding and time-consuming, requiring a significant input of measured data in order to predict relevant optical parameters correctly~\cite{Liu10, Tuk10}.

An improved description of multiple light scattering can be obtained by analytical theories originating from nanophotonics, wherein multiple scattering of light is described from first principles~\cite{Lag96, Ros99, Bre05, Akk07}.  The advantage of such \textit{ab initio} models over ray tracing is that one obtains fundamental physical insight, starting from the detailed nanostructure of a sample.  Moreover, calculating optical properties from a model is computationally much faster than performing a ray tracing simulation.

Recently, the transport mean free path was obtained for white LED diffusion plates by applying diffusion theory to diffuse transmission measurements~\cite{Vos13}.  To interpret the results, an \textit{ab initio} model was made without adjustable parameters by using Mie theory.  While the model was found to agree well with the experimental results, both the experiment and the model did not include absorption or emission, both of which are quintessential in the functioning of a white LED.  The aim of our present paper is to close this gap and understand multiple scattering in a functional phosphor.

We examine the competitive interplay of multiple scattering and absorption, and elucidate the underlying physics of light transport.  The roles of multiple light scattering and energy conversion are highly intertwined; in this paper we show how we can disentangle them with experimental measurements of diffuse intensity.  We present the mean free paths of scattering and absorption, a set of characteristic diffusion parameters that encapsulates the underlying mechanisms from first principles.
We compare their values to the thickness of the LEDs' phosphor plate, which is an essential parameter in multiple light scattering~\cite{Lag96, Ros99, Bre05}.
This represents the first time white LEDs have been studied with a combined experimental and analytical approach from first principles based on diffusion theory.  Our approach produces explicit physical parameters that allow for a non-phenomenological, physically motivated treatment.  As such, this approach offers concrete physical insights, allowing comparison of phosphors of even vastly different structural and optical properties, such as particle shape or size distribution.

\section{Light diffusion with energy conversion - observables and optical properties}

Multiple light scattering in diffusive photonic media is often studied by total transmission $T$ - also known as diffuse transmission - which is the transmission integrated over all outgoing angles at which light exits from a medium.  In general the total transmission contains information on the transport mean free path $\ell_{tr}$, a crucial parameter that describes multiple scattering of light~\cite{Lag96, Ros99, Bre05, Akk07, Dur94, Vos13}. The transport mean free path is the distance it takes for the direction of light to become randomized while light performs a random walk in a scattering medium.

The measurement technique we have used is similar at first glance to a measurement of total transmission.  We have made a measurement of the angularly integrated light intensity at the exit face of our phosphor plates.  However in addition to multiple scattering, there is also energy transfer in a phosphor plate: light leaving the plate consists not only of light that is diffusely transmitted, but also of light that has been absorbed from one spectral band and re-emitted into another band.  Hence the result of integrating the outgoing light intensity is a complex dispersive sum of total transmission, absorption, and emission.  From here on we shall refer to this measured quantity of total transmission with energy transfer as the \textit{total relative intensity} $T_{rel}$

\begin{equation}\label{eq:tridef}
T_{rel}(\lambda) \equiv \frac{I_{tot}(\lambda)}{I_{0}(\lambda)}
\end{equation}

\noindent where $I_{tot}$ is the angular integrated light intensity at the exit face of the phosphor plate, and $I_{0}$ is the reference intensity.  In our study the reference intensity is chosen to be the integrated incident intensity in the absence of the phosphor plate.

We can separate the outgoing intensity ratio into two constituent terms:

\begin{equation}\label{eq:tridef2}
T_{rel}(\lambda) = T(\lambda) + T_{em}(\lambda) =  \frac{I(\lambda)}{I_{0}(\lambda)} + \frac{I_{em}(\lambda)}{I_{0}(\lambda)}.
\end{equation}

\noindent $T$ is the total transmission in the presence of absorption and $T_{em}$ is the ratio of re-emitted intensity to the reference intensity.  We can lump multiple scattering with the first step of energy conversion, namely absorption, into one term $T$, because in diffusion theory the combined behavior can be readily described simply as lossy diffusive scattering~\cite{Gar92}.  Thus the total transmission $T$ contains information on both the transport mean free path $\ell_{tr}$ and the absorption mean free path $\ell_{abs}$.  The absorption mean free path is the distance it takes for light to be absorbed to a fraction $1/e$ while light performs a random walk in a scattering medium.  We note that $T$ remains an integrated measure of the incident light that is \textit{transmitted} through the phosphor plate.

 In the limit where there is negligible absorption, we define $T(\ell_{tr}, 1/\ell_{abs} \rightarrow 0) \equiv T^{0}(\ell_{tr})$.  In phosphor, $T^{0}(\ell_{tr})$ is an experimentally accessible quantity if we remove the effects of absorption and emission by spectral filtering. From $T^{0}(\ell_{tr})$ the transport mean free path can be deduced, as was studied in Ref.~\cite{Vos13}.  Once $\ell_{tr}(\lambda)$ is known, we can use this information to deduce $\ell_{abs}(\lambda)$ from $T$.  We assume there is no re-absorption, which is reasonable as the Stokes-shifted light emitted by the phosphor will no longer be in the phosphor absorption band. Therefore by a suitable combination of experiments in different wavelength ranges with excitation filters, we can distinguish the multiple light scattering from diffuse absorption properties in a white LED phosphor.

\section{Experimental details}

We present results obtained on polymer plates containing phosphorescent material, typical for white LED units of the Fortimo type~\cite{For13}.  The ceramic phosphor YAG:Ce$^{3+}$ is the most commonly-used compound for remote phosphors and is also employed in Fortimo white LEDs.  We investigated five polymer plates (polycarbonate, Lexan 143R) containing YAG:Ce$^{3+}$ crystalline particles with weight densities ranging through $\phi = 2.0, 2.5, 3.0, 3.5$ and $4.0$~wt\%, the typical densities of phosphor used in Fortimo white LEDs.  These phosphor particles are known to have a broad size distribution centered around 10~$\mu$m.  The emission spectrum of the particles in powder form was measured at Philips Lighting.  Plates were fabricated as follows: a powder of YAG:Ce$^{3+}$ crystals is mixed with polymer by making a compound in the required weight ratio.  Next, the compound is shaped into plates (60~mm x 2~mm) by industrial injection molding.  The plates have a thickness $L = 2.0$~mm.

To measure broadband, spectrally-resolved, total relative intensity $T_{rel}$ of light we implemented a setup with a white-light source, which in our case is a white LED (Luxeon LXHL-MW1D). The spectral range runs from $\lambda = 400$~nm to $700$~nm as shown in Fig.~\ref{fig:fullTT}(b). The output of such a white LED light source has a spectral range identical to the range of interest, conveniently eliminating the need for additional spectral filters.  To measure the total transmission in the absence of absorption $T^{0}$, we use a broadband high-pass filter with a cut-off wavelength at $520$~nm that removes the blue component of the white-light source.

The beam was collimated and the diameter was set by several irises to 2 mm, and was incident at normal incidence on a plate that was vertically placed in front of the entrance port of an integrating sphere.
Since the incident beam was not focused, we can safely neglect the dependence of total transmission on the angle of incidence.
We verified that the entrance port of the integrating sphere was sufficiently large to accept the complete diffuse spot emanating from the strongest scattering sample.
The diffuse output of the integrating sphere was monitored with a fiber-to-chip spectrometer (Avantes) with a spectral resolution of $1.2$~nm.

For all plates, we collected several spectra and repositioned a sample in between measurements to test the reproducibility and sample homogeneity; the measurements reproduced to within a few percent on different days.
To calibrate the total relative intensity or total transmission values we collected reference spectra $I_{0}$ in absence of a sample, and determined the total relative intensity $T_{rel}(\lambda)$ or the total transmission $T(\lambda)$ as the ratio of a sample spectrum and a reference spectrum.
Reference spectra were frequently collected in between sample spectrum  measurements to correct for possible time-dependent changes in the setup.
The total relative error in the transmission is estimated to be about $5 \%$ percent points, including systematic errors, based on the variations between different measurements.

\section{Results and discussion}

\subsection{Total transmission}

\begin{figure}[htbp]
\centerline{\includegraphics[width=.8\columnwidth]{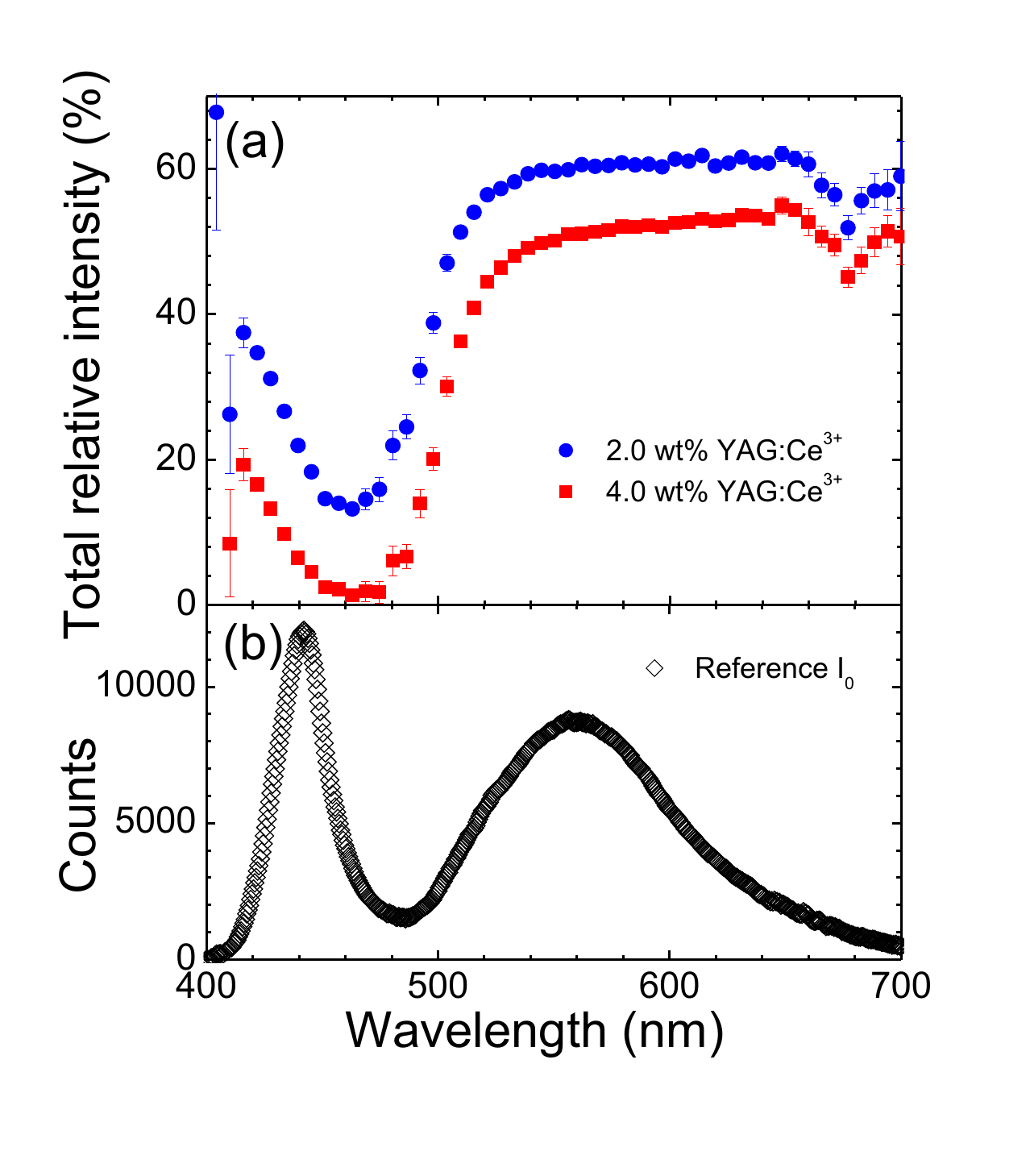}}
\caption{(a) Total relative intensity $T_{rel}(\lambda)$ versus wavelength for polymer plates with a phosphor density of 2 and 4 wt\% YAG:Ce$^{3+}$.  (b) The spectrum of the reference intensity $I_{0}$.}
\label{fig:fullTT}
\end{figure}

In Figure~\ref{fig:fullTT}, we show total relative intensity spectra $T_{rel}(\lambda)$ for phosphor plates with increasing phosphor content.  For visual clarity we have only plotted $T_{rel}(\lambda)$ of the lowest (2\%) and highest (4\%) weight percentages studied.  Also for visual clarity we have binned $T_{rel}(\lambda)$ into 6~nm-wide spectral intervals.  The spectra are limited to the range 400 to 700~nm since this is the range where the white light source intensity $I_{0}$ is significant, see Fig.~\ref{fig:fullTT}(b).
In each spectrum, $T_{rel}(\lambda)$ exhibits a broad trough between $\lambda =$ 400 and 510~nm, indicating a strong absorption of blue light by the phosphor.

\begin{figure}[htbp]
\centerline{\includegraphics[width=.8\columnwidth]{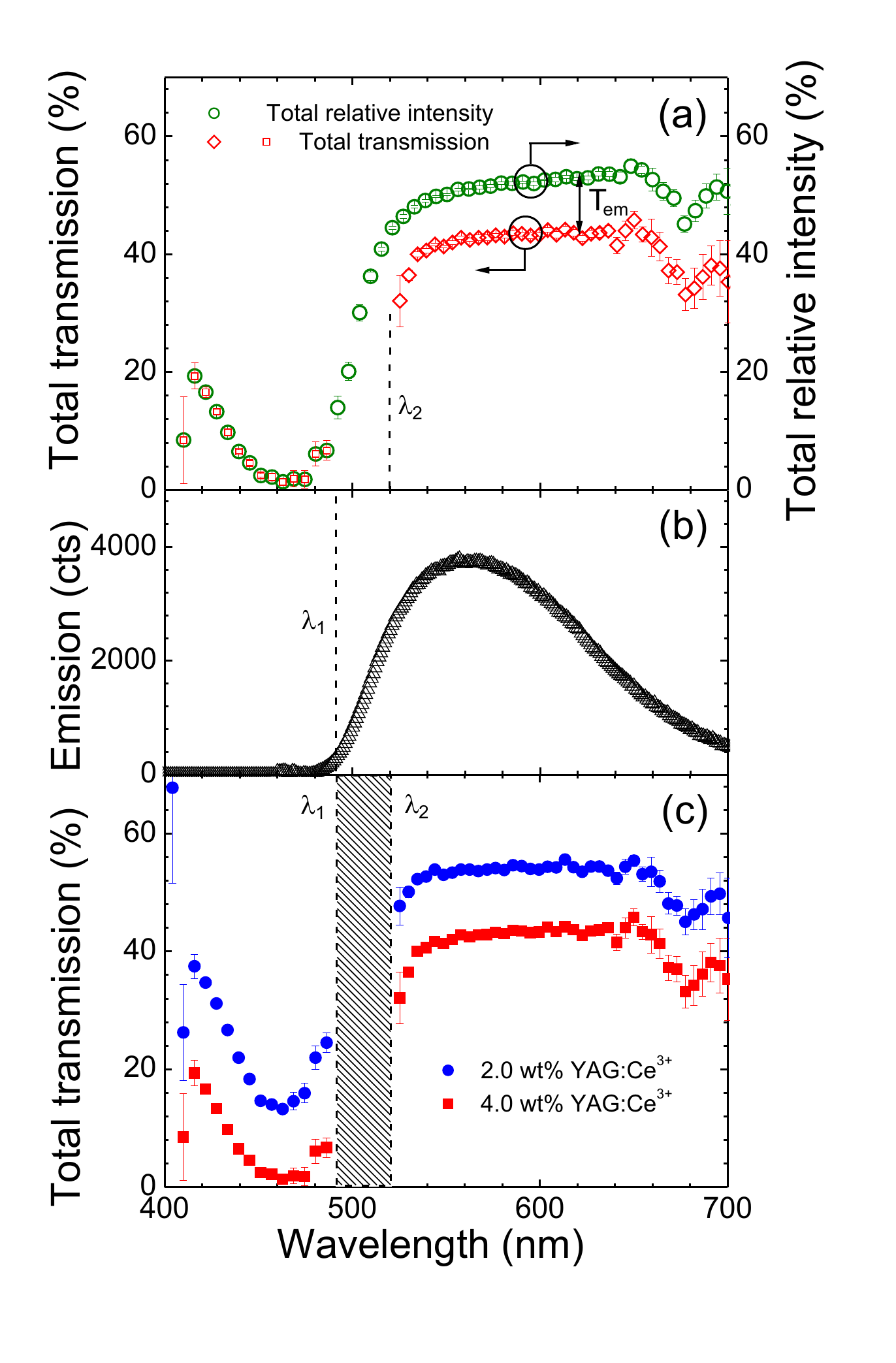}}
\caption{(Color online) (a) Total relative intensity $T_{rel}(\lambda)$ versus wavelength for a phosphor density of 4 wt\% YAG:Ce$^{3+}$.  The original data from Fig.~\ref{fig:fullTT} are indicated as green triangles.  The total transmission $T(\lambda)$ obtained from a measurement with a high-pass filter is shown as red diamonds.  $\lambda_{2}$ indicates the wavelength cut-off of the filter.  (b) The phosphor emission spectrum of YAG:Ce$^{3+}$. $\lambda_{1}$ is the wavelength where we estimate the onset of emission. (c) The total transmission $T(\lambda)$ for a phosphor density of 2 and 4 wt\% YAG:Ce$^{3+}$.  The spectral region from 490 to 540~nm is excluded due to the presence of emission and absorption simultaneously.}
\label{fig:truncTT}
\end{figure}

We know that due to the phosphorescence of YAG:Ce$^{3+}$, $T_{rel}(\lambda)$ includes light absorption and re-emission.  To distinguish the contributions, we first measured the spectral range of YAG:Ce$^{3+}$ emission, shown in Figure~\ref{fig:truncTT}(b).  We see that the emission has an onset at $\lambda_{1} = 490$~nm and extends to all longer wavelengths.
Therefore we can already conclude with Eq.~\ref{eq:tridef2} that for short wavelengths $\lambda \leq \lambda_{1}$, the total transmission $T$ is equal to the total relative intensity: $T(\lambda \leq \lambda_{1}) = T_{rel}(\lambda \leq \lambda_{1})$.
The equivalence of $T_{rel}(\lambda \leq \lambda_{1})$ and $T(\lambda \leq \lambda_{1})$ is illustrated in Figure~\ref{fig:truncTT}(a) by the overlap of the red and green symbols.

To identify the total transmission at long wavelengths in the range of the phosphor emission, we performed experiments with an optical filter with a cut-off wavelength at $\lambda_{2} = 520$~nm.  Such a filter prevents blue light from exciting the phosphor, and thus corresponds to zero emitted intensity: $T_{em} = 0$.  Hence at long wavelengths $\lambda > \lambda_{2}$ ($= 520$~nm) the total transmission is equal to the total relative intensity: $T(\lambda > \lambda_{2}) = T_{rel}(\lambda > \lambda_{2})$.
The total relative intensity in the presence of the filter is shown as red diamonds in Figure~\ref{fig:truncTT}(a) for the 4~wt\% YAG:Ce$^{3+}$ plate.
Thus by combining the measurements of $T(\lambda)$ for $\lambda < \lambda_{1}$ and $\lambda > \lambda_{2}$, we obtain the total transmission of a phosphor-containing diffuser plate for the entire visible spectral range, with the exception of the 30-nm interval between $\lambda_{1}$ and $\lambda_{2}$ as is shown in Figure~\ref{fig:truncTT}(a).\footnote{We have chosen $\lambda_{2}$ so that there is some overlap with the YAG:Ce$^{3+}$ absorption band; in this way we monitor the onset of absorption in our measurement of $T(\lambda)$.}

Comparing the two curves $T_{rel}(\lambda)$ and $T(\lambda)$ in Fig.~\ref{fig:truncTT}(a), we see that $T_{rel}(\lambda)$ is significantly higher than $T(\lambda)$ in the wavelength range $520 < \lambda < 700$~nm as a result of emitted light.  Apparently, the shape of the emitted spectrum $I_{em}(\lambda)$ agrees well with the input spectrum of the white source $I_{0}(\lambda)$, resulting in a nearly wavelength-independent $T_{em}(\lambda)$ shown in Figure~\ref{fig:truncTT}(a).
We conclude that since the relative total intensity for $\lambda > \lambda_{2}$ in Fig.~\ref{fig:fullTT} is enhanced by the phosphor emission, only measuring the relative total intensity does not allow us to characterize light scattering in phosphors with emission.

In Figure~\ref{fig:truncTT}(c), we compare total transmission spectra for samples with a YAG:Ce$^{3+}$ content of 2~wt\% and 4~wt\%.  We observe a decreasing $T(\lambda)$ with increasing phosphor density.  In the range $\lambda > \lambda_{2}$ this behavior is intuitively reasonable as the scattering strength is expected to increase with the density of the phosphor particles that act as scatterers.
In the range $\lambda < \lambda_{1}$ ($= 490$~nm) the total transmission is affected by absorption of light in the YAG:Ce$^{3+}$ phosphor; as is evident from the spectral shape with a minimum at $\lambda = 460$~nm.

\subsection{Transport mean free path}

From diffusion theory, the analytical solution of the diffusion equation in the presence of absorption is known \cite{Gar92,Gar08}.  The total transmission $T$ is equal to:

\begin{equation} \label{eq:fullsoln}
 T = (1 - R_{s})\frac{1}{\alpha z_{e}} \frac{\mathrm{sinh}[\alpha (z_{p} + z_{e})] \mathrm{sinh}(\alpha z_{e})}{\mathrm{sinh}[\alpha (L + 2z_{e})]}
\end{equation}

\noindent The total transmission $T(L, \lambda)$ is a function of the thickness of the plate $L$ and the wavelength $\lambda$.  $R_{s}(\lambda)$ is the specular reflectivity of the incident light from the front surface.  As the phosphor density is relatively low, the specular reflectivity is mainly determined by the refractive index of the polymer matrix.  Previous measurements show that in the visible regime, the specular reflectivity is low, between 4 and 5\%~\cite{Vos13}.  Therefore in the present work we can safely neglect the specular reflectivity.  The penetration depth $z_{p}$ is the distance within the medium at which a source of isotropically emitted light is taken to be positioned, and is $2/3 \ell_{tr}$ for diffuse incident light and $\ell_{tr}$ for collimated normal-incident light.  The extrapolation length $z_{e}$ is the distance outside the boundary at which the diffuse intensity decreases to zero~\cite{Ish78} and it equals
\begin{equation} \label{eq:extrapol}
z_{e} = \frac{1}{2 \alpha} \mathrm{ln} \left( \frac{1 + \alpha z_{0}}{1 - \alpha z_{0}} \right).
\end{equation}

\noindent The reciprocal absorption mean free path $\alpha$ is equal to $\alpha = 1/\ell_{abs}$, where $\ell_{abs}$ is the diffuse absorption length.  In the absence of absorption the extrapolation length $z_{0}$ is well-described by
\begin{equation} \label{eq:extrapol2}
z_{0} = \frac{2}{3} \ell_{tr} \left( \frac{1 + \bar{R}}{1 - \bar{R}} \right),
\end{equation}

\noindent where $\bar{R}$ is the angle- and polarization-averaged reflectivity~\cite{Lag89}.  For a medium with an average refractive index of $n =$~1.5 typical of a polymer the average diffuse reflectivity is equal to $\bar{R} =$~0.57~\cite{Zhu91}.

From the total transmission in Eq.(\ref{eq:fullsoln}), we wish to determine firstly the transport mean free path $\ell_{tr}$ and secondly the absorption mean free path $\ell_{abs}$ as a function of wavelength $\lambda$ and phosphor density. Using the Taylor series expansion for $\mathrm{sinh}(x)$, we first write Eq.(\ref{eq:fullsoln}) in the limit of low absorption ($\alpha << 1$) as
\begin{equation}
T_{tr, \alpha \rightarrow 0} = (1 - R_{s})\frac{1}{\alpha z_{e}} \frac{\alpha (z_{p} + z_{e}) (\alpha z_{e})}{\alpha (L + 2z_{e})}.
\end{equation}

\noindent As a result, Eq.(\ref{eq:fullsoln}) converges to the standard expression for the total transmission $T_{tr}^{0}$~\cite{Dur94}:
\begin{equation} \label{eq:noabsorp}
T_{tr}^{0} =  (1 - R_{s})\frac{z_{p} + z_{0}}{L + 2z_{0}}
\end{equation}

\noindent where we take $z_{p} = \ell_{tr}$, as the incident light is collimated.

\begin{figure}[htbp]
\centerline{\includegraphics[width=.75\columnwidth]{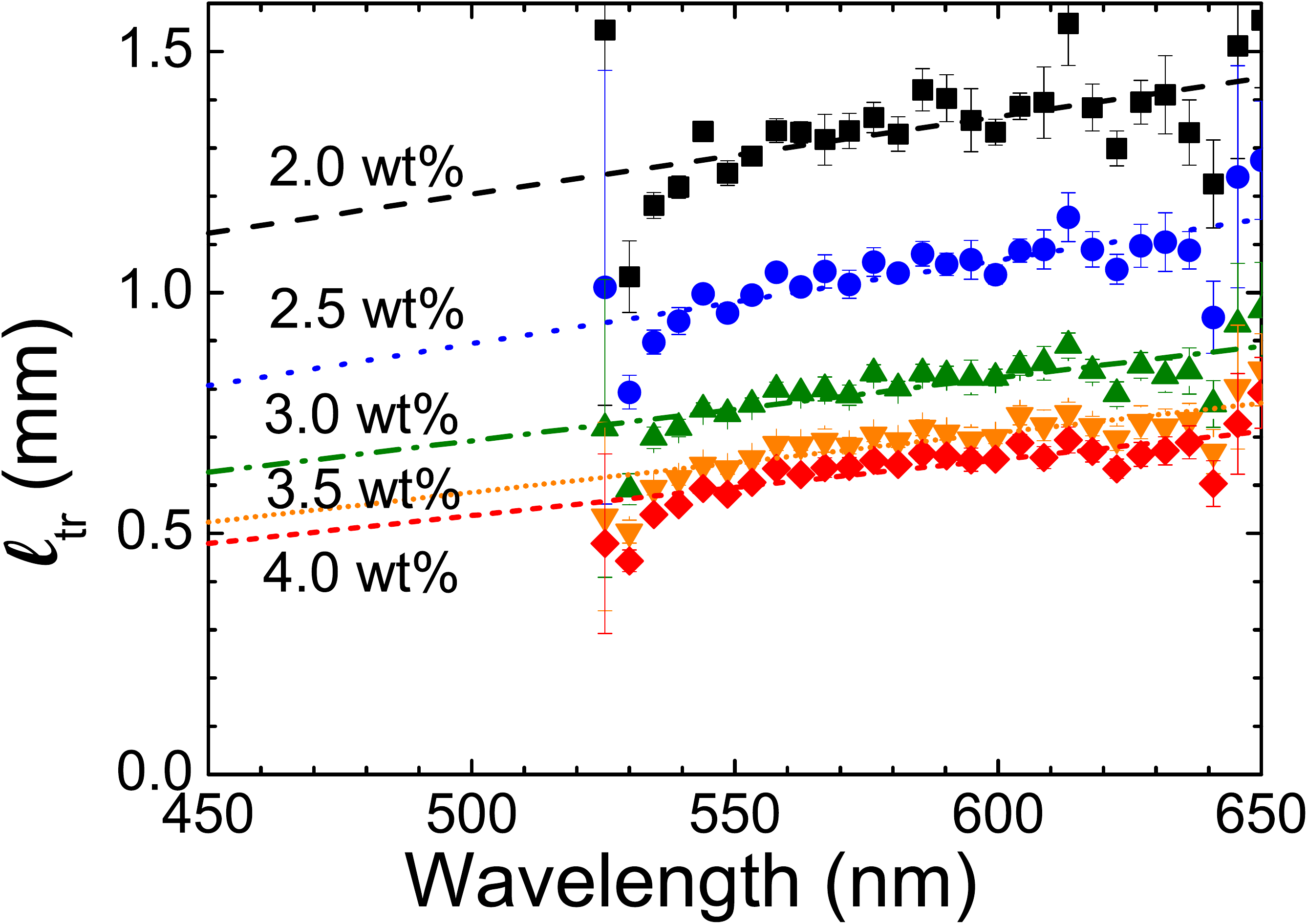}}
\caption{(Color online) Transport mean free path $\ell_{tr}$ versus wavelength in the range of weak absorption, extracted through Eq.~\eqref{eq:noabsorp} for different phosphor densities (symbols).  By taking $\ell_{tr}$ as a linear function of $\lambda$, we obtain by extrapolation $\ell_{tr}(\lambda < \lambda_{2})$ from the measured data shown in Fig.~\ref{fig:truncTT} (lines).}
\label{fig:ltrspectra}
\end{figure}

From the total transmission results in the absence of absorption in Fig.~\ref{fig:truncTT} ($\lambda > \lambda_{2}$), the transport mean free path can be found by applying Eq.~\eqref{eq:noabsorp}. Plotting  $\ell_{tr}$ as a function of $\lambda$ (Fig.~\ref{fig:ltrspectra}), we see firstly that the mean free path decreases with increasing phosphor density.  Thus, scattering increases with increasing density, as expected.
We also see that there is generally more scattering for blue than red light, a trend consistent with that observed for ensembles of highly polydisperse non-absorbing nano-particle scatterers~\cite{Vos13}.  It has been shown that in highly polydisperse media, the transport mean free path $\ell_{tr}$ is only weakly dependent on wavelength~\cite{Riv99,Mus08}.  We can thus well describe the behavior of $\ell_{tr}$ as a linear function of $\lambda$ from 450 to 700~nm, the most important spectral range for white LEDs . Such a linearity was also observed in the study of LED diffuser plates with highly polydisperse ensembles of  scatterers reported in Ref.~\cite{Vos13}.  Therefore to obtain Fig.~\ref{fig:ltrspectra}, the total transmission is linearly extrapolated into the absorption region of $\lambda < 530$~nm, and diffusion theory is used to find the transport mean free path $\ell_{tr}$.  For the strongest scattering sample of 4 wt\% YAG:Ce$^{3+}$, the mean free path increases by a factor 1.4 in the spectral range between 450 and 650~nm.  This reveals that the observed scattering is not in the Rayleigh regime, which has a $\lambda^{4}$ scaling of the scattering cross section that would correspond to an increase by a factor of 4~\cite{Hul57,Boh83}.

\begin{figure}[htbp]
\centerline{\includegraphics[width=.75\columnwidth]{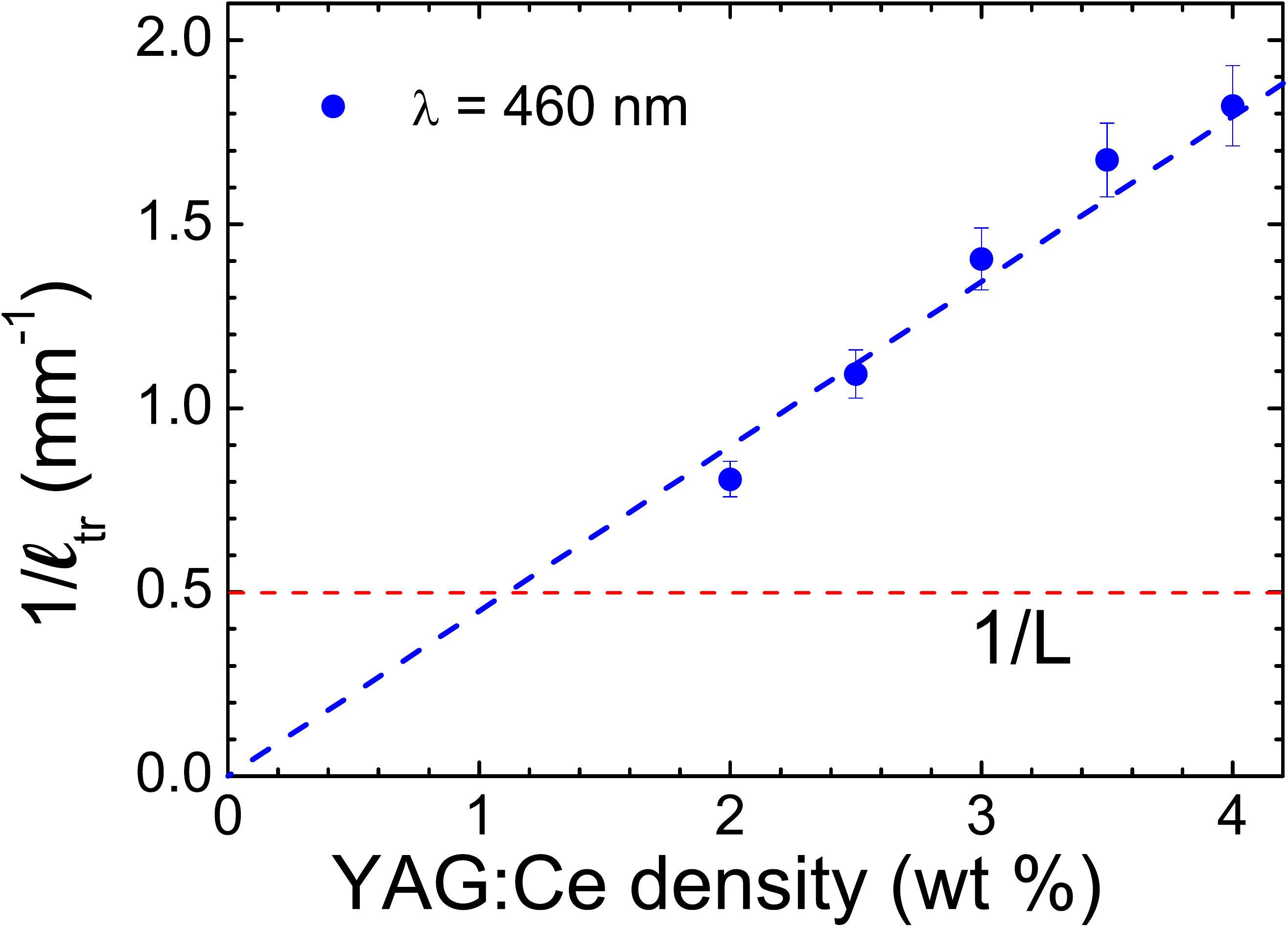}}
\caption{Inverse of the transport mean free path $\ell_{tr}$ versus phosphor density at $\lambda = 460$~nm.  The line has a slopes of 0.45 $\pm$ 0.01~(mm wt$\%)^{-1}$.  The reciprocal sample thickness $1/L$ is indicated by the dashed red line.}
\label{fig:inverseltr}
\end{figure}

Fig.~\ref{fig:inverseltr} shows the effect of increasing the density of phosphor on the scattering properties.  To this end we have plotted the inverse transport mean free path $\ell_{tr}^{-1}$ for phosphor densities between 2 to 4\%.  We compare $\ell_{tr}^{-1}$ for a wavelength $\lambda =$ 460~nm at the absorption maximum of phosphor. First of all, we see that the sample thickness is 2 to 3 times that of the transport mean free path.  We conclude that light is completely randomized as it traverses the thickness of the plate, and that the densities of phosphor optimized for use in white LED components are in a regime of moderate multiple scattering.  Secondly, we find that $\ell_{tr}^{-1}$ has a linear dependence on phosphor density, \textit{i.e.}, phosphor is the main contributor to scattering in our samples. This is reasonable as our samples contain no added scatterers.

Since the mean free path of light in our samples are comparable to the sample thickness ($\ell \simeq L$), in other words the samples are moderately in the multiple scattering regime, it is relevant to wonder if it is justified to invoke diffusion theory.
In Ref.~\cite{Mar88}, the enhanced backscattering cone, a quintessential multiple scattering phenomenon, was calculated both with exact theory and with diffusion theory. Even for low-order of scattering, \emph{i.e.}, for weak multiple scattering, the diffusion theory agrees very well - better than $10 \%$ - with the exact results. In addition, simulations of the diffuse transmission of samples with widely varying thickness show that the diffusion approximation is in surprisingly good agreement with the exact results, even for a sample thickness of the order of the mean free path~\cite{Dur94}. Therefore, we conclude that is justified to apply diffusion theory to our samples.

\begin{figure}[htbp]
\centerline{\includegraphics[width=.75\columnwidth]{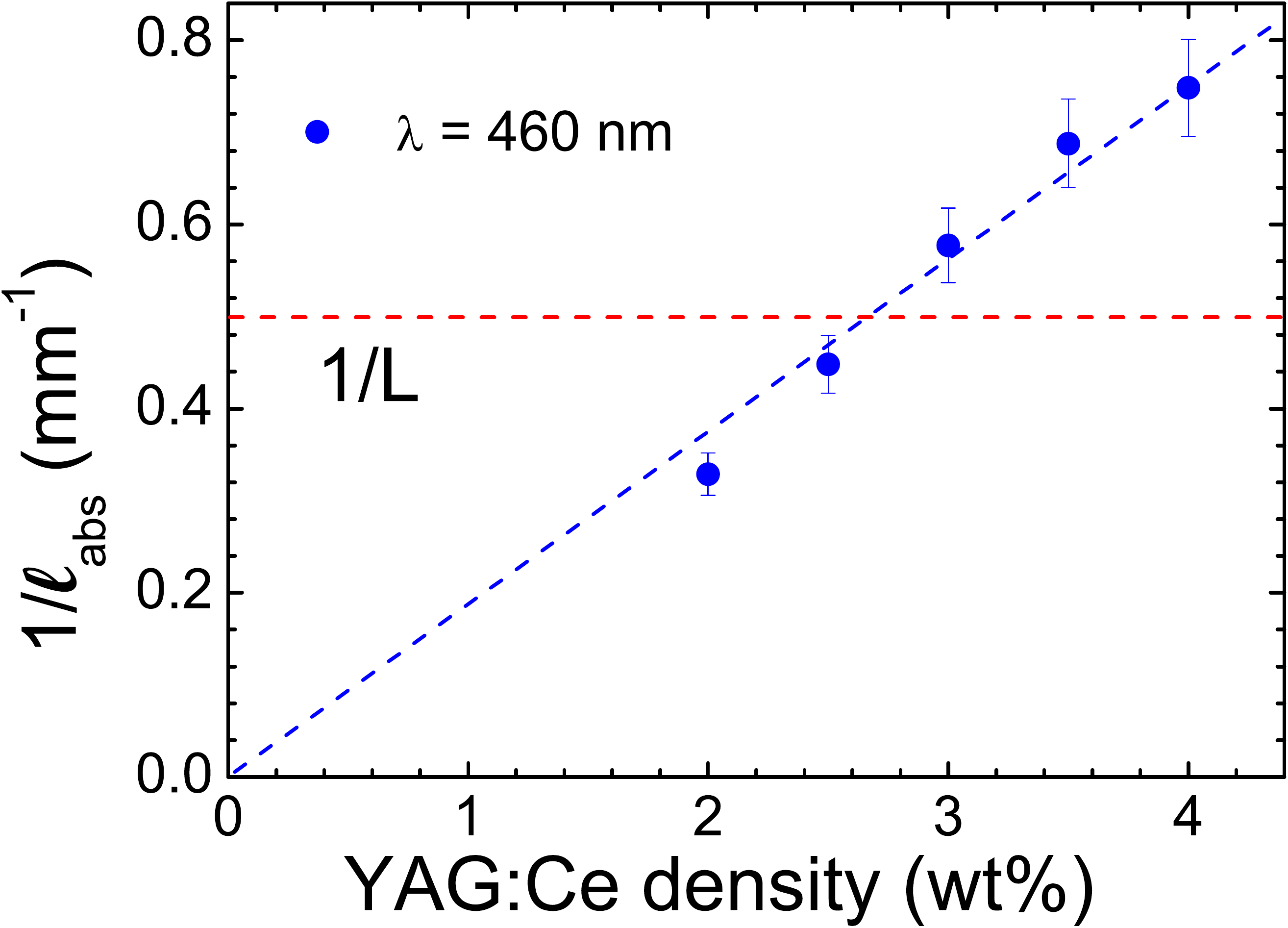}}
\caption{Inverse of the absorption mean free path $\ell_{abs}$ versus YAG:Ce$^{3+}$ density at $\lambda = 460$~nm. The line has a slope of 0.19 $\pm$ 0.01~(mm wt$\%)^{-1}$.  The reciprocal sample thickness $1/L$ is indicated by the dashed red line.}
\label{fig:inverselab}
\end{figure}

\subsection{Absorption mean free path}

To find the absorption mean free path of phosphor, we use the analytical expression for the total transmission with absorption Eq.~\eqref{eq:fullsoln}.  We know $\ell_{tr}$ and $T(\lambda)$ for $\lambda < \lambda_{2} = 530$~nm, and all other parameters in Eq.~\eqref{eq:fullsoln} are also known.  Therefore we can deduce the absorption mean free path $\ell_{abs}$, using a procedure outlined in detail in the Appendix.  The resulting inverse absorption lengths as a function of phosphor density are shown in Fig.~\ref{fig:inverselab}.  We see that $\ell_{abs}^{-1}$ increases linearly with phosphor density.  This linearity confirms our expectation that phosphor is the main absorbing component in the sample.  Comparing Figures~\ref{fig:inverseltr} and \ref{fig:inverselab}, we see that for a given phosphor density the transport mean free path is less than the absorption mean free path, with a relation $\ell_{abs} \approx 2.5 \ell_{tr}$.
This implies that the phosphor particles scatter light more strongly than they absorb light.
This might seem counter-intuitive, as we normally think of phosphor only in terms of its function in color conversion.
Our results show that physically, phosphor is in fact first and foremost a \textit{scattering} material.

\begin{figure}[htbp]
\centerline{\includegraphics[width=.75\columnwidth]{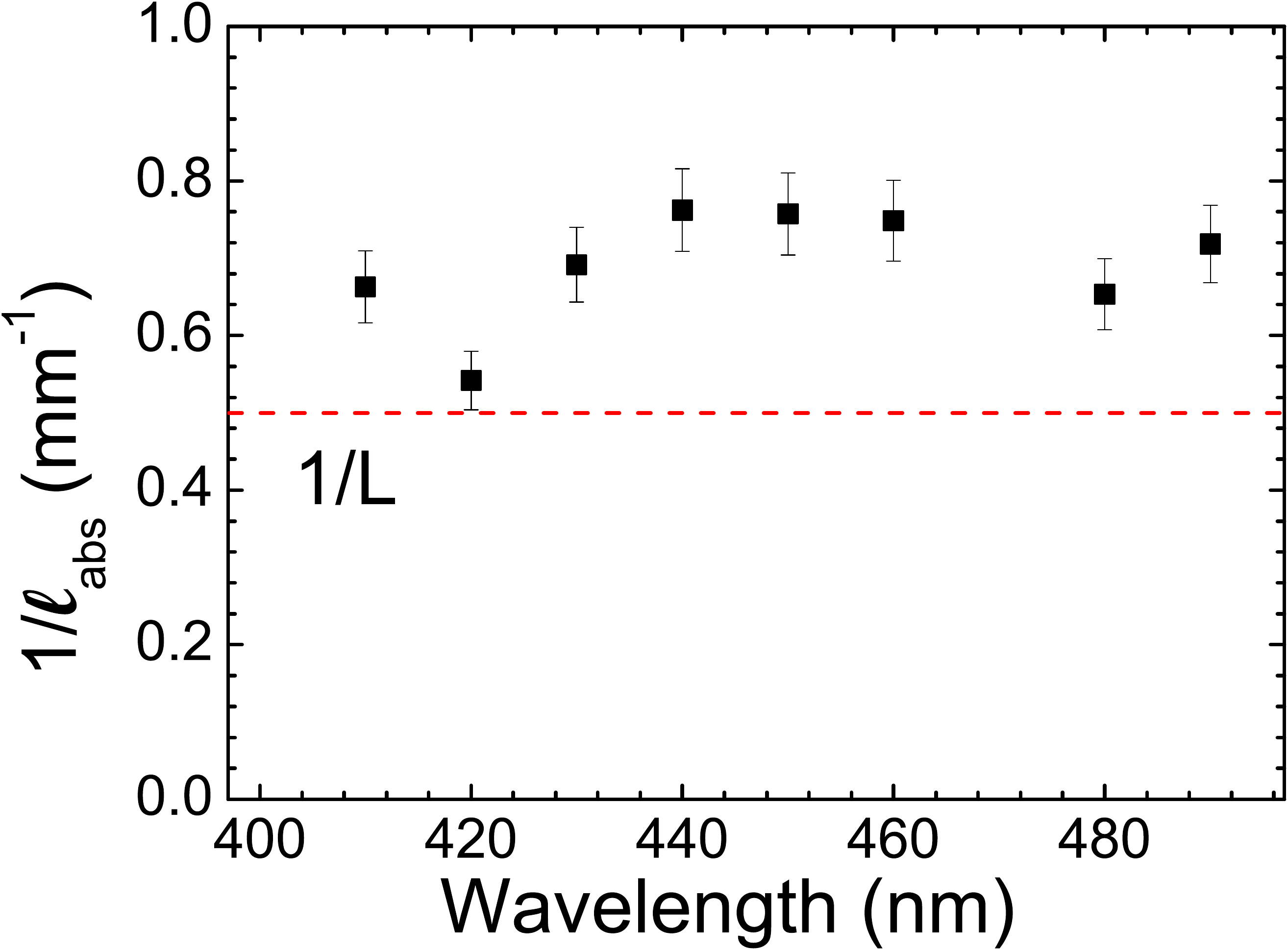}}
\caption{Spectral distribution of the inverse absorption mean free path $1/\ell_{abs}$ for a phosphor plate containing 4 wt\% YAG:Ce$^{3+}$.  The reciprocal sample thickness $1/L$ is indicated by the dashed red line.}
\label{fig:spectrallab}
\end{figure}

Fig.~\ref{fig:spectrallab} shows the wavelength-dependence of the inverse absorption mean free path for the sample containing 4 wt\% YAG:Ce$^{3+}$.  Both Figs.~\ref{fig:inverselab} and ~\ref{fig:spectrallab} indicate that across the absorption band, $\ell_{abs} \simeq L$.  From Fig.~\ref{fig:inverselab} we can clearly see that the optimal phosphor densities arrived at by trial-and-error coincides with a value of $\ell_{abs}$ that ranges from being slightly less to slightly greater than the sample thickness $L$. This is intuitively sensible, as luminaires are optimized for the generation of diffuse blue light and an optimal, but not total, conversion to yellow light.  Therefore the optimal diffuse white illumination is best accomplished for $\ell_{abs}$ comparable to $L$.

\begin{figure}[htbp]
\centerline{\includegraphics[width=.75\columnwidth]{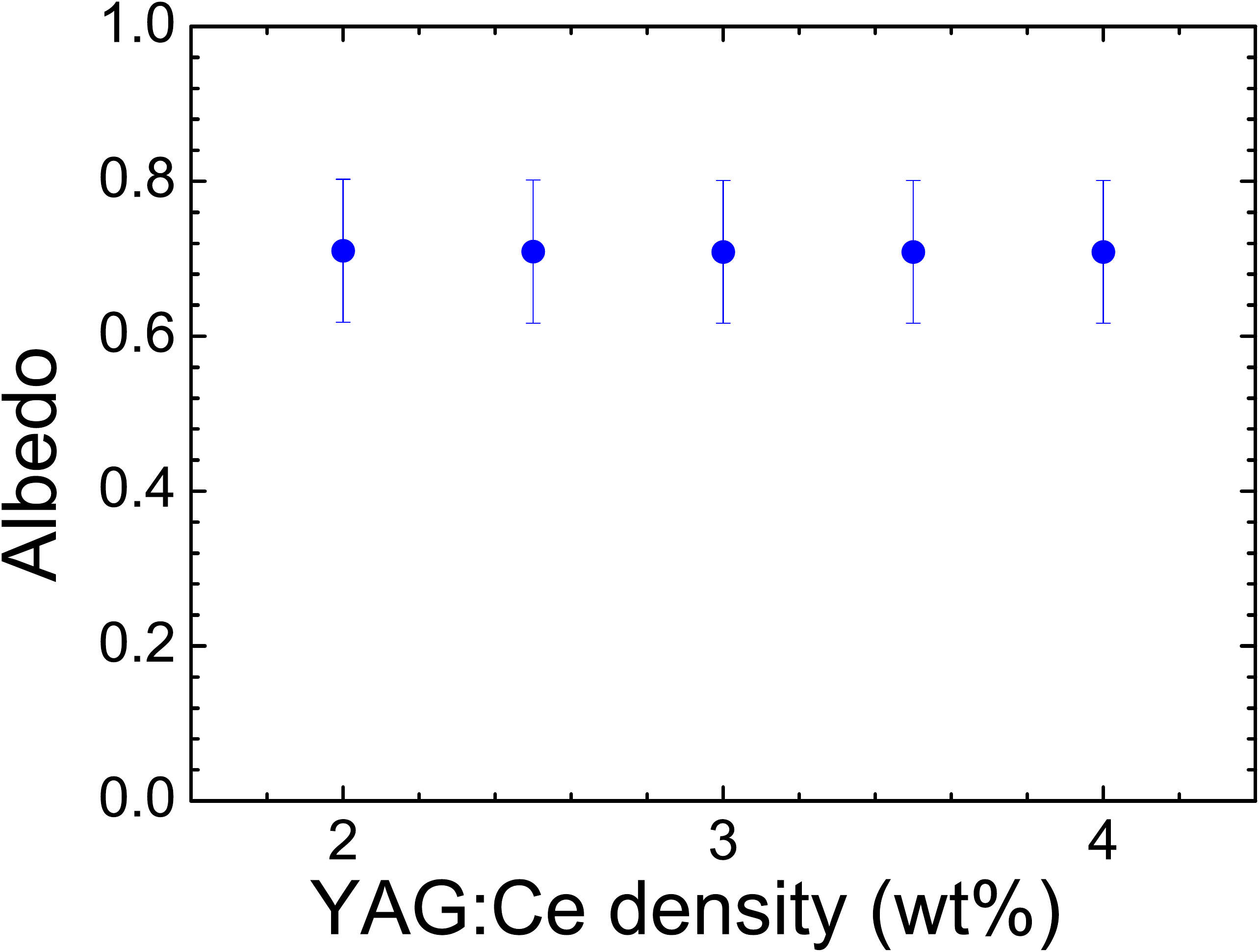}}
\caption{The albedo $a$ at a wavelength $\lambda = 460$~nm as a function of phosphor density.}
\label{fig:albedo}
\end{figure}

\begin{figure}[htbp]
\centerline{\includegraphics[width=.75\columnwidth]{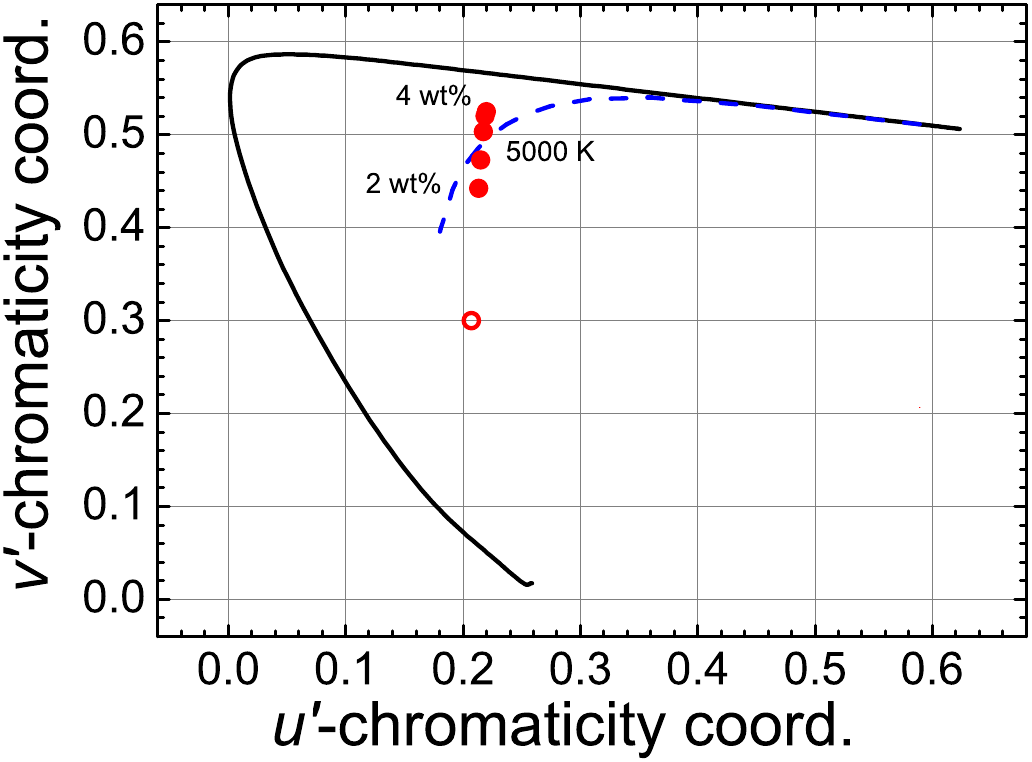}}
\caption{Color coordinates of the phosphor plates with increasing phosphor density on the CIE 1960 $u', v'$ chromaticity diagram (filled red circles).  The plates were illuminated by a reference light source with $u'_{i} = 0.21, v'_{i} = 0.30$ (open red circle).  The solid black line denotes the boundary of the $u', v'$ color space.  The dashed blue line denotes the Planckian locus. Combining phosphor plates of increasing density with the reference light source creates a color point sequence which intersects the Planckian locus at a color temperature of 5000~K. }
\label{fig:colorpoint}
\end{figure}

The interplay between scattering and absorption in phosphor is concisely described by the albedo $a$ that is defined as $a = \sigma_{sc}/(\sigma_{sc} + \sigma_{abs})$, where $\sigma_{sc}$ and $\sigma_{abs}$ are the scattering and absorption cross-sections~\cite{Ish78}.
We find that the albedo is 0.7 $\pm$ 0.1 for $\lambda = 460$~nm.  This result is independent of phosphor density, as we would expect in the case where scattering and absorption are single-particle properties (Fig.~\ref{fig:albedo}).  The albedo allows us to place the optical behavior of phosphor doped polymers in relation to other complex photonic media.  Traditionally, investigations of multiple light scattering have focused on systems with a high albedo of $a > 0.98$.  A very low albedo, as observed here, is rarely encountered in light scattering, typically in systems of truly black particles~\cite{Boh83}.
The results we present here thus represent one of the few systems studied with low albedo.  Besides the phosphor component of white LEDs, possible applications where this kind of low-albedo system is interesting include paint, printing~\cite{Meu12}, and solar fuel cells~\cite{Nag10}.

Since low albedo systems seem to be rarely studied, it is relevant to wonder how well systems with low albedo are described by diffusion theory.  This was investigated by van der Mark \textit{et al.} in the context of describing the scattering intensity in enhanced backscattering experiments from strongly scattering media~\cite{Mar88}.  The authors found that the bistatic scattering coefficient $\gamma$ calculated by diffusion theory agrees remarkably well with exact calculations even for $a = 0.6$.  Therefore we conclude that it is justified to apply diffusion theory to our samples.

Finally let us place our approach in the context of the uniform chromaticity diagram commonly used in optical engineering (CIE, 1960)~\cite{Mal11}. Our results can be translated into the $u', v'$ coordinates in this color space, as shown in Fig.~\ref{fig:colorpoint}.  Spectral data of a reference light source is used as the initial starting point $u'_{i} = 0.21, v'_{i} = 0.30$.  This is convolved with our total transmission data to give new final values $u'_{f}, v'_{f}$ as a function of phosphor density.  In this way, we effectively describe a white LED with a remote phosphor in a simple way.
Increasing phosphor  density increases $u'_{f}, v'_{f}$.  For our choice of reference light source, increasing the density of phosphor in the phosphor plates creates a color point sequence which crosses the Planckian locus or black-body line at a color temperature of 5000~K.
The color temperature of a white LED is of utmost importance to its application.  Our work establishes a physics-based connection between a white LED with a certain $u', v'$ color point, and its displacement in this color space with the addition of scatterers or absorbers.
By disentangling the roles of absorption and scattering into fundamental physical parameters based on diffusion theory, we now have a way to predict shifts in $u'_{f}, v'_{f}$ due to changes in $\ell_{tr}$ and $\ell_{abs}$.  This offers a practical method to model changes in chromaticity in the phosphor of a white LED from a relatively straightforward measurement of the total transmission, and will be considered further in future research.

\section{Summary and outlook}

We have studied the optical properties of phosphorescent plates of YAG:Ce$^{3+}$ in a polycarbonate matrix, a widely employed production method for white LED modules.   We measured the total relative intensity over the visible wavelength range, and extracted from this measurement the total transmission.  Employing photonic diffusion theory, we obtained from the total transmission the transport mean free path.  The transport mean free path, which fundamentally characterizes the diffuse transport of light, is studied as a function of phosphor density. In addition we have experimentally differentiated the optical roles of phosphor in multiple scattering and absorption.  This enabled us to obtain the absorption mean free path which characterizes the color conversion of light in the phosphor plates.

We find that the densities of phosphor optimized for use in white LED components exhibit transport and absorption mean free paths which are comparable to each other and also to the thickness of the sample.  The optimized system is thus a system with weak scattering and low albedo.  Our study opens the way to a purely physical description of color conversion in white LEDs based on diffusion.  Our method is based on experimental and analytical tools that lead to a direct interpretation in terms of translational displacements in $u', v'$ color space.

The application of our method is two-fold: as a stand-alone physical description with predictive power, and as a complement to numerical methods.  Diffusion equation problems with analytical solutions are usually limited to simple sample geometries, such as a plate, a sphere, or a semi-infinite medium.  Understanding of diffusion in these geometries will already provide physical insights with predictive potential, which could guide the evolution of efficient phosphor layers with improved optical properties. To model complex LED luminaires, we foresee the complementary use of our diffusion approach to augment ray-tracing techniques by providing physically motivated input parameters.  Complementary use of our method to improve numerical simulations reduces the time demands currently required in measuring supplementary data in order to predict optical properties accurately.

\section{Appendix: Finding the absorption mean free path}

To deduce the absorption mean free path in the spectral region of significant absorption $\lambda < \lambda_{2} = 530$~nm, we begin with the analytical solution for the total transmission Eq.~\eqref{eq:fullsoln}.  As a result of the analysis of Fig.~\ref{fig:truncTT} and the extrapolation of the transport mean free path into the absorption region, we know $\ell_{tr}(\lambda < \lambda_{2})$.  From the initial experimental data presented in Fig.~\ref{fig:fullTT}, we also know $T(\lambda < \lambda_{2})$. Thus we are left with only one unknown quantity in Eq.~\eqref{eq:fullsoln}: the absorption mean free path $\ell_{abs}(\lambda < \lambda_{2})$.

\begin{figure}[htbp]
\centerline{\includegraphics[width=.75\columnwidth]{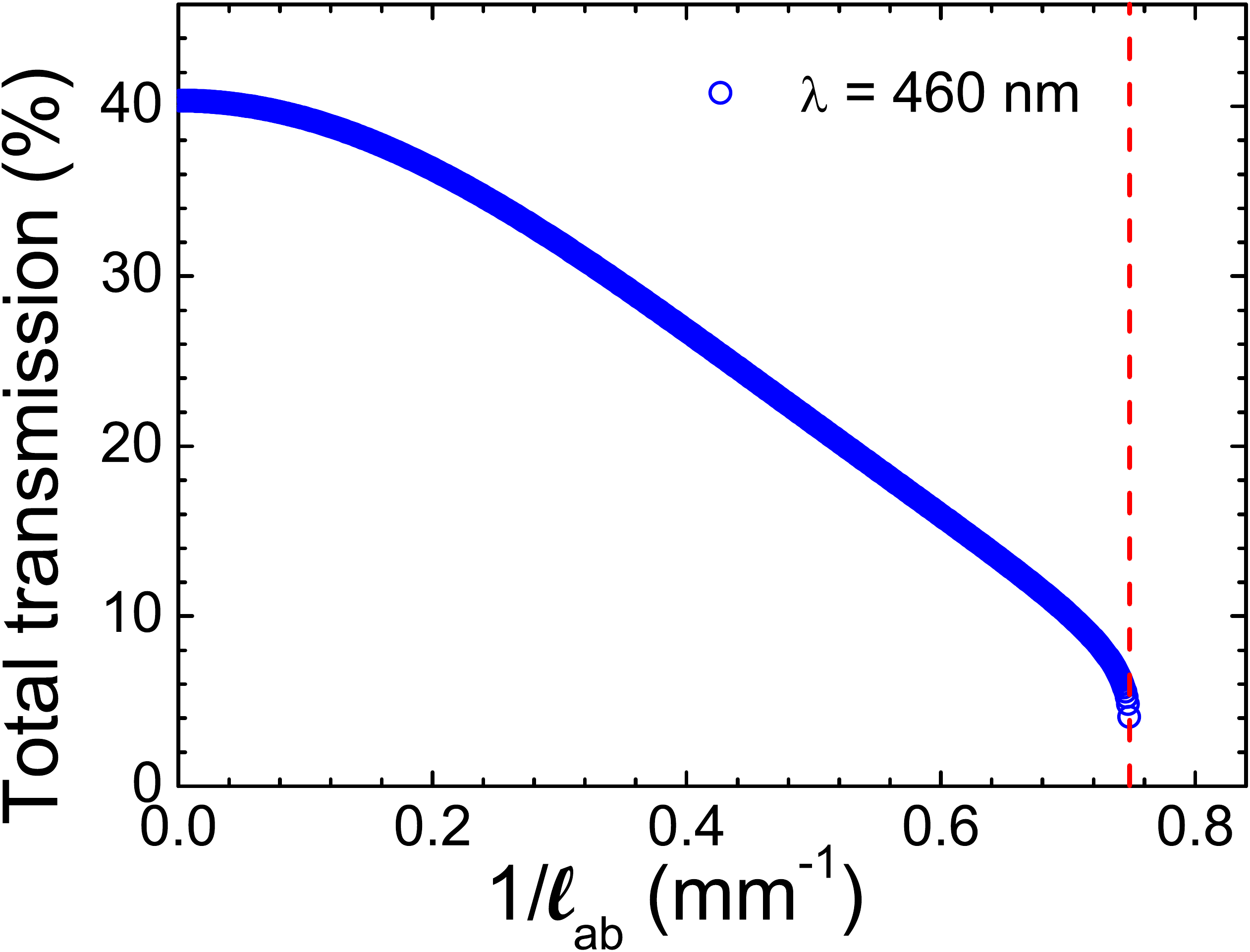}}
\caption{Total transmission versus inverse absorption mean free path $1/\ell_{abs}$ for a polymer plate with 4 wt\% YAG:Ce$^{3+}$ density, at $\lambda = 460$~nm, within the absorption region of YAG:Ce$^{3+}$.  Calculated from Eq.~\eqref{eq:fullsoln} for a value of $\ell_{tr}$ extracted from the total transmission measurements shown in Fig.~\ref{fig:truncTT}. The red dashed line indicate the asymptotic value of the inverse absorption mean free path $1/\ell_{abs}$ for strong absorption.}
\label{fig:calc}
\end{figure}

Fig.~\ref{fig:calc} is a plot of Eq.~\eqref{eq:fullsoln} as a function of the total transmission $T$ and the inverse absorption mean free path $\ell_{abs}^{-1}$.  In Fig.~\ref{fig:calc} we have fixed the transport mean free path in Eq.~\eqref{eq:fullsoln} to its experimental value at peak absorption, $\ell_{tr}(\lambda = 460)$~nm. We can therefore find the specific inverse absorption mean free path that corresponds to the experimentally obtained total transmission $T(\lambda = 460)$~nm (Fig.~\ref{fig:fullTT}).  This procedure for extracting the absorption mean free path from total transmission data allows us to obtain inverse absorption mean free paths as a function of phosphor density, plotted in Fig.~\ref{fig:inverselab}.

It can be seen from Fig.~\ref{fig:calc} that in Eq.~\eqref{eq:fullsoln}, $T(L, \lambda)$ asymptotically approaches a limiting value of $\ell_{abs}^{-1}$.
The asymptotic behavior of $T(L, \lambda)$ stems from the extrapolation length in the presence of absorption $z_{e}$, given in Eq.~\eqref{eq:extrapol}.  The quantity $z_{e}$ diverges if the absorption mean free path $\ell_{abs}$ becomes smaller than the extrapolation length without absorption $z_{0}$.  We can see that in Eq~\eqref{eq:extrapol2}, when $1 - \alpha z_{0} > 1$, the logarithmic part of Eq.~\eqref{eq:extrapol} becomes divergent.  In other words, Eq.~\eqref{eq:fullsoln} is unphysical for very strong absorption compared to scattering, or $\ell_{tr} >> \ell_{abs}$.  For our samples we do not reach this regime; as we have shown, the optimal balance of scattering and absorption in white LED phosphors is when the two mean free paths are nearly commensurate.  In our case, the asymptotic behavior of $T(L, \lambda)$ results in a small variation of $\ell_{abs}^{-1}$ with $\lambda$ at regions of high absorption such as at $\lambda = 460$~nm, the absorption peak of our data.  The broadband spectral dependence of the inverse absorption mean free path for $\lambda < \lambda_{2}$ is shown in Fig.~\ref{fig:spectrallab}.

\section{Acknowledgments}

VL would like to thank Cock Harteveld for technical advice, and Jochen Aulbach, Jacopo Bertolotti, Timmo van der Beek for discussions.  This work was supported by the Dutch Technology Foundation STW (contract no. 11985), and by FOM and NWO, and the ERC.


\end{document}